\documentclass[twocolumn,showpacs,preprintnumbers,amsmath,amssymb,amsfonts,superscriptaddress]{revtex4}

\usepackage{graphicx}

\newcommand{\be}{\begin{equation}}
\newcommand{\beq}{\begin{equation}}
\newcommand{\en}{\end{equation}}
\newcommand{\eeq}{\end{equation}}
\newcommand{\bea}{\begin{eqnarray}}
\newcommand{\ena}{\end{eqnarray}}
\newcommand{\hbo}{\hbox to 1 true cm {\hfill } }

\newcommand{\GeV}{{\rm GeV}}

\newcommand{\Os}{\mathcal{O}_{\rm S}}
\newcommand{\Og}{\mathcal{O}_{\rm G}}

\begin{document}

\preprint{KEK Preprint 2013-9, Edinburgh 2013/11}

\title{Light composite scalar in twelve--flavor QCD on the lattice}

\author{Yasumichi~Aoki}
\author{Tatsumi~Aoyama}
\author{Masafumi~Kurachi}
\author{Toshihide~Maskawa}
\author{Kei-ichi~Nagai}
\author{Hiroshi~Ohki}
\affiliation{Kobayashi-Maskawa Institute for the Origin of Particles and the Universe (KMI), Nagoya University, Nagoya, 464-8602, Japan}
\author{Enrico Rinaldi}
\affiliation{Kobayashi-Maskawa Institute for the Origin of Particles and the Universe (KMI), Nagoya University, Nagoya, 464-8602, Japan}
\affiliation{Higgs Centre for Theoretical Physics, SUPA, School of Physics and Astronomy, University of Edinburgh, Edinburgh EH9 3JZ, UK}
\author{Akihiro~Shibata}
\affiliation{Computing Research Center, High Energy Accelerator Research Organization (KEK), Tsukuba 305-0801, Japan}
\author{Koichi~Yamawaki}
\author{Takeshi~Yamazaki}
\affiliation{Kobayashi-Maskawa Institute for the Origin of Particles and the Universe (KMI), Nagoya University, Nagoya, 464-8602, Japan}
\collaboration{LatKMI collaboration}
\noaffiliation

\date{
\today
}

\begin{abstract}
Based on lattice simulations using highly improved staggered quarks for twelve--flavor QCD with several bare fermion masses, we observe a flavor--singlet scalar state lighter than the pion in the correlators of fermionic interpolating operators. The same state is also investigated using correlators of gluonic interpolating operators. Combined with our previous study, that showed twelve--flavor QCD to be consistent with being in the conformal window, we infer that the lightness of the scalar state is due to infrared conformality. This result shed some light on the possibility of a light composite Higgs boson (``technidilaton'') in walking technicolor theories.
\end{abstract}

\pacs{12.38.Gc,  11.15.Ha,  12.39.Mk,  12.60.Nz}
\keywords{ technicolor, lattice, glueballs, composite higgs }

\maketitle

At the Large Hadron Collider (LHC), the existence of a new bosonic particle of mass $m_H \approx 125\ \GeV$~\cite{:2012gk,:2012gu}, identified as the Higgs boson, has recently been confirmed. Although recent analyses show its consistency with the Standard Model Higgs, a possibility still allowed from current data is that this boson is a composite particle, coming from a new high--energy strongly--interacting gauge theory. A typical example is the walking technicolor theory~\cite{Yamawaki:1985zg} (see also~\cite{Holdom:1984sk, Akiba:1985rr,Appelquist:1986an}), featuring approximate scale invariance and a large anomalous dimension, $\gamma_m \approx 1$. Such a theory predicts a light composite Higgs, ``technidilaton''~\cite{Yamawaki:1985zg}, a light scalar particle emerging as a (pseudo) Nambu--Goldstone boson of the spontaneously broken approximate scale symmetry. In this context, we consider one family of theories, the SU($3$) gauge theory with $N_f$ massless Dirac fermions in the fundamental representation (dubbed large $N_f$ QCD). Large $N_f$ QCD has been studied by many groups using different lattice discretizations and techniques in search for a candidate of walking technicolor (for reviews, see~\cite{DelDebbio:2011rc,Neil:2012cb,Giedt:2012hg}).

We (LatKMI collaboration) put effort in studying $N_f=4,8,12,16$ QCD using lattice simulations with a common setup. Studies of the conformal and walking properties of the $N_f=12$ and $8$ theories have already been published~\cite{Aoki:2012eq,Aoki:2013xza}. Here, we would like to focus on the $N_f=12$ theory which we found~\cite{Aoki:2012eq} to be consistent with the conformal theory, and having unbroken chiral symmetry.  Pressed and inspired by the impressive experimental discovery of the Higgs boson, we present our results for the mass of the lightest scalar bound state in the $N_f=12$ SU($3$) gauge theory using fully non--perturbative lattice Monte Carlo simulations. This measurement of the non--perturbative scalar spectrum is important because it gives informations on the possible conformal dynamics at low energy and it serves as a model towards study of the composite Higgs in the walking technicolor, if not a candidate as it stands for the walking technicolor. As we will explain in the following, the scalar spectrum is incredibly challenging and those challenges are added to the usual problems faced by lattice calculations near the conformal window~\cite{DelDebbio:2011rc,Neil:2012cb,Giedt:2012hg}.

In this Letter, we report the results of our calculations, which show a scalar state lighter than $\pi$ ($N_f$--flavor extension of the pion). A preliminary report on such a light scalar was given in Ref.~\cite{Aoki:2013pca}. In QCD, the lightest flavor--singlet scalar state is the $f_0(500)$($\sigma$) meson, whose mass has been reviewed in the latest Particle Data Group~\cite{Beringer:1900zz}. Several studies of the $\sigma$ meson were carried out in lattice QCD~\cite{PhysRevD.65.014508,McNeile:2000xx,Kunihiro:2003yj,Hart:2006ps,Bernard:2007qf,Fu:2012gf}. Another example of a flavor--singlet scalar particle would be the $0^{++}$ glueball, whose existence as a resonance in QCD has yet to be proven (see Ref.~\cite{Ochs:2013fk} for a detailed review). Two--pions scattering states could also be relevant to this channel when combined in an $s$--wave~\cite{Morningstar:2013bda}. In QCD, we also expect a mixing between gluonic and fermionic degrees of freedom and this could be the case also for larger $N_f$. Note that the candidate for a composite Higgs or a technidilaton must be predominantly a fermionic bound state and not a glueball state, since the gluons do not carry SU($2$) $\times$ U($1$) charges. Our contribution is the first of this kind for a theory that could be relevant for beyond the Standard Model physics.

To study flavor--singlet states using lattice simulations, the computation of disconnected diagrams is mandatory for a correct estimate of their mass. This requires computationally expensive measurements and high statistics in order to give results with relatively small errors. Previous studies of the scalar spectrum using fermionic operators in $N_f=12$ QCD either did not include the computation of disconnected diagrams~\cite{Fodor:2011tu}, or were restricted to an unphysical region of the parameter space that is not related to the continuum limit physics of the asymptotically free theory~\cite{Jin:2012dw}.

We discretize the continuum SU($3$) gauge theory with $12$ degenerate fermions using $3$ degenerate staggered fermion species of bare mass $m_f$ (each coming in $4$ tastes). In this letter all the dimensionful quantities are expressed in lattice units. At finite lattice spacing, where the simulations take place, the continuum flavor symmetry does not hold exactly. We use a tree--level Symanzik gauge action and the highly improved staggered quark (HISQ)~\cite{Follana:2006rc} action without the tadpole improvement 
and the mass correction in the Naik term~\cite{Bazavov:2011nk} for the fermions. The flavor symmetry breaking of this action is highly suppressed in QCD~\cite{Bazavov:2011nk} and we observed that it is almost negligible in our $N_f = 12$ 
QCD simulations~\cite{Aoki:2012eq}. At fixed lattice spacing, defined by the bare coupling constant $\beta=6/g^2=4.0$, we simulate three physical volumes $L^3$ with $L=24, 30, 36$ and aspect ratio $T/L = 4/3$. We investigate the flavor--singlet scalar spectrum at four different bare quark masses $m_f = 0.05, 0.06, 0.08,$ and $0.10$. These parameters allow us to check for finite size systematics and to test hyperscaling~\cite{Miransky:1998dh,DelDebbio:2010jy}.

We carry out the simulations by using the standard hybrid Monte-Carlo (HMC) algorithm using MILC code version 7~\cite{MilcCode} with some modifications to suit our needs, such as the Hasenbusch mass preconditioning~\cite{Hasenbusch:2001ne} to reduce the large computational cost at the smaller $m_f$. Beside the excellent flavor(--taste) symmetry, another important feature of our simulations is the large number of Monte Carlo trajectories from uninterrupted Markov chains obtained after more than $1000$ trajectories for thermalization. For all sets of parameters explored, we collect between $8000$ and $30000$ trajectories and we do measurements every $2$ trajectories. This is a necessary step to contrast the rapid degradation of the signal in the flavor--singlet scalar correlators. The simulation parameters and number of trajectories for each parameter are tabulated in Table~\ref{tab:simulations}. For the measurement of the ground state mass of this channel we used interpolating operators including both the fermionic fields and the gauge fields, with the appropriate quantum numbers.
The statistical errors for the fermionic and gluonic measurements are estimated by jackknife method with bin size of $200$ and $160$ trajectories, respectively.

 \begin{table}
  \begin{tabular}{c|c|r|l|l|l}
     $L^3 \times T$ & $m_f$ & \multicolumn{1}{|c}{$N_{\rm cfgs}$} & \multicolumn{1}{|c|}{$m_\sigma$} & \multicolumn{1}{c}{$m_\pi$} & \multicolumn{1}{|c}{$m_\sigma/m_\pi$}\\
     \hline
     $24^3 \times 32$ & 0.05 &11000 & 0.237(13)($^{02}_{01}$) & 0.3273(19)$^*$ & 0.73(4)($^{1}_{0}$) \\
     $24^3 \times 32$ & 0.06 & 14000 & 0.279(17)($^{07}_{01}$) & 0.3646(16)$^*$ & 0.77(5)($^{2}_{0}$) \\
     $24^3 \times 32$ & 0.08 & 15000 & 0.359(21)($^{01}_{18}$) & 0.4459(11)& 0.81(5)($^{0}_{4}$) \\
     $24^3 \times 32$ & 0.10 &  9000 & 0.453(42)($^{37}_{08}$) & 0.5210(7) & 0.87(8)($^{7}_{2}$) \\
     $30^3 \times 40$ & 0.05 & 10000 & 0.275(13)($^{21}_{08}$) &  0.3192(14)$^*$ & 0.86(4)($^{7}_{3}$)\\
     $30^3 \times 40$ & 0.06 & 15000 & 0.329(15)($^{47}_{12}$) & 0.3648(9)$^*$ & 0.90(4)($^{13}_{3}$)\\
     $30^3 \times 40$ & 0.08 & 15000 & 0.382(21)($^{03}_{16}$) & 0.4499(8) & 0.85(5)($^{1}_{4}$)\\
     $30^3 \times 40$ & 0.10 &  4000 & 0.431(51)($^{06}_{04}$) & 0.5243(7) & 0.82(10)($^{1}_{1}$)\\
     $36^3 \times 48$ & 0.05 &  5000 & 0.283(23)($^{01}_{02}$) & 0.3204(7)$^*$ & 0.88(7)($^{0}_{1}$)\\
     $36^3 \times 48$ & 0.06 &  6000 & 0.305(22)($^{25}_{06}$) & 0.3636(9)$^*$ & 0.84(6)($^{7}_{2}$)\\
   \end{tabular}
   \caption{\label{tab:simulations} Parameters of lattice simulations for $N_f=12$ QCD at fixed $\beta=4.0$. $N_{\rm cfgs}$ is the number of saved gauge configurations. The second error of $m_\sigma$ is a systematic error coming from the fit range. The values of $m_\pi$ are from Ref.~\cite{Aoki:2012eq}, but the ones with ($^*$) have been updated. The error on $m_\sigma/m_\pi$ comes only from $m_\sigma$.}
 \end{table}


In our fermionic scalar calculation, we employ the local fermionic bilinear operator 
\begin{equation}
  \label{eq:fermionic-scalar-op}
  \Os(t) \; = \; \sum_{i=1}^{3}\sum_{\vec{x}} \overline{\chi}_i(\vec{x},t) \chi_i(\vec{x},t) \ ,
\end{equation}
where the index $i$ runs through different staggered fermion species. The explicit staggered spin--taste structure of the bilinear operator can be written as $\overline{\chi}_i(y+A) ({\bf 1} \otimes {\bf 1})_{AB} \chi_i(y+B)$ with $y$ as an origin of the hypercube, and $A$, $B$ as vectors in the hypercube. Note that this system has exact symmetry for exchanging the species. The taste symmetry breaking, which is to vanish in the continuum limit, is very small in our simulations. Therefore, a part of the full flavor symmetry is exact, and the rest is only broken by a small amount. From $\Os(t)$ we calculate the correlator, which is constructed by both the connected $C(t)$ and vacuum--subtracted disconnected $D(t)$ correlators, $\langle \Os(t) \Os^\dag(0) \rangle = 3 D(t) - C(t)$, where the factor in front of $D(t)$ comes from the number of species. It is noted that the contribution of $D(t)$ with respect to $C(t)$ increases with $N_f = $ \#species$\times 4$.

The operator $\Os$ overlaps with the flavor--singlet scalar state ($\sigma$), but also with a flavor non--singlet pseudo--scalar state (${\pi_{\rm \overline{SC}}}$), which is the staggered parity partner of $\sigma$; therefore, in the large--time limit, the correlator above behaves as
\begin{equation}
  \label{eq:fermionic-combined}
3 D(t) - C(t) = A_\sigma(t) + (-1)^t A_{\pi_{\rm \overline{SC}}}(t) \ ,
\end{equation}
where $A_H(t) = A_H (e^{-m_H t} + e^{-m_H (T-t)})$, and the pseudo--scalar state has a $(\gamma_5\gamma_4 \otimes \xi_5\xi_4)$ spin--taste structure, but is species--singlet. 

Because $C(t)$ can be regarded as a flavor non--singlet scalar correlator, it should have a contribution from the lightest non--singlet scalar state ($a_0$) (e.g. $a_0(980)$ in QCD~\cite{Beringer:1900zz}), and its staggered parity partner ($\pi_{\rm SC}$). When $t$ is large, we can therefore write
\begin{equation}
  \label{eq:fermionic-connected}
 -C(t) \; = \; A_{a_0}(t) + (-1)^t A_{\pi_{\rm SC}}(t) \ ,
\end{equation}
where both $a_0$ and $\pi_{\rm SC}$ are species non-singlet and have the same taste structure as $\sigma$ and $\pi_{\rm \overline{SC}}$, respectively. The $\pi_{\rm SC}$ state is degenerate with the  $(\gamma_5 \otimes \xi_5)$ $\pi$ and also with $\pi_{\rm \overline{SC}}$ ($m_{\pi_{\rm SC}} = m_{\pi} = m_{\pi_{\rm \overline{SC}}}$) when the taste symmetry, thus the full flavor symmetry, is recovered.

\begin{figure}[ht]
  \includegraphics*[height=5cm]{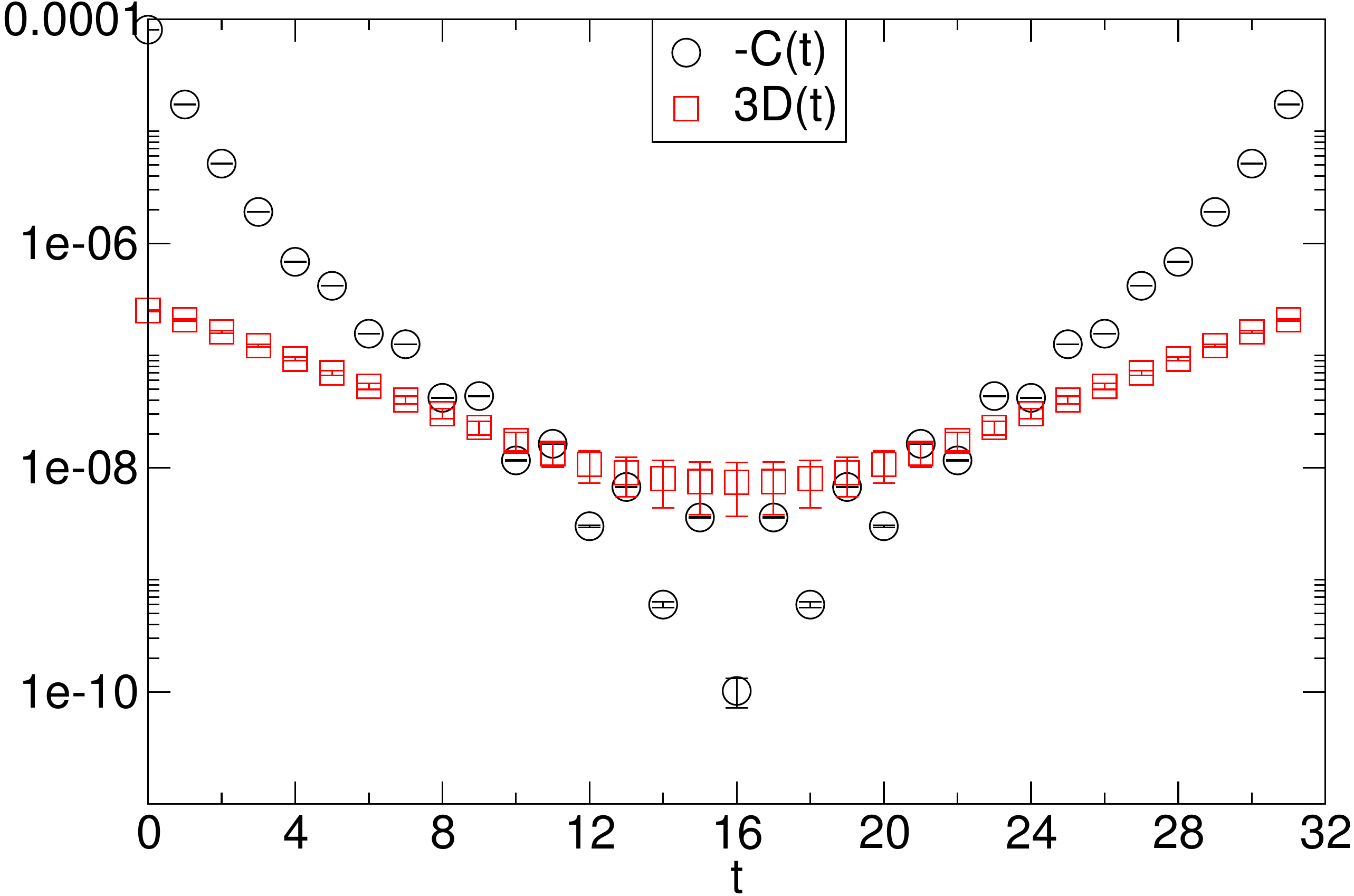}
  \caption{\label{fig:propagators} Connected $-C(t)$ and disconnected $3D(t)$ correlators for $L=24$ and $m_f=0.06$.}
\end{figure}

The disconnected correlator $D(t)$, which is essential to obtain the $\sigma$ mass, can be calculated by inverting the staggered Dirac operator at each space--time point $(\vec{x},t)$. The computational cost of this inversion is mitigated by using a stochastic noise method. Moreover, its large fluctuations 
from the random noise in the method is dealt with by using a variance reduction method already employed for the flavor--singlet pseudo--scalar~\cite{Venkataraman:1997xi,Gregory:2007ev} and chiral condensate~\cite{McNeile:2012xh} in usual QCD, and for the flavor--singlet scalar meson in $N_f = 12$ QCD~\cite{Jin:2012dw}. We employ $64$ spacetime random sources for this reduction method. From Eq.~\eqref{eq:fermionic-combined} and Eq.~\eqref{eq:fermionic-connected}, the large--time asymptotic form of $3D(t)$ can be written as 
\begin{equation}
  \label{eq:fermionic-disconnected-only}
  3 D(t) \; = \; A_{\sigma}(t) - A_{a_0}(t) + (-1)^t
(A_{\pi_{\rm SC}}(t) - A_{\pi_{\rm \overline{SC}}}(t))\ .
\end{equation}
A typical result for $-C(t)$ and $3D(t)$ is shown in Fig.~\ref{fig:propagators}. In the large--time region, $3D(t)$ behaves as a smooth function of $t$ in contrast to $-C(t)$, which has a clear oscillating behavior. This means that the taste--symmetry breaking between $A_{\pi_{\rm SC}}(t)$ and  $A_{\pi_{\rm \overline{SC}}}(t)$ in Eq.~\eqref{eq:fermionic-disconnected-only} is small, as expected from our previous work~\cite{Aoki:2012eq}.

\begin{figure}[ht]
  \includegraphics*[height=5cm]{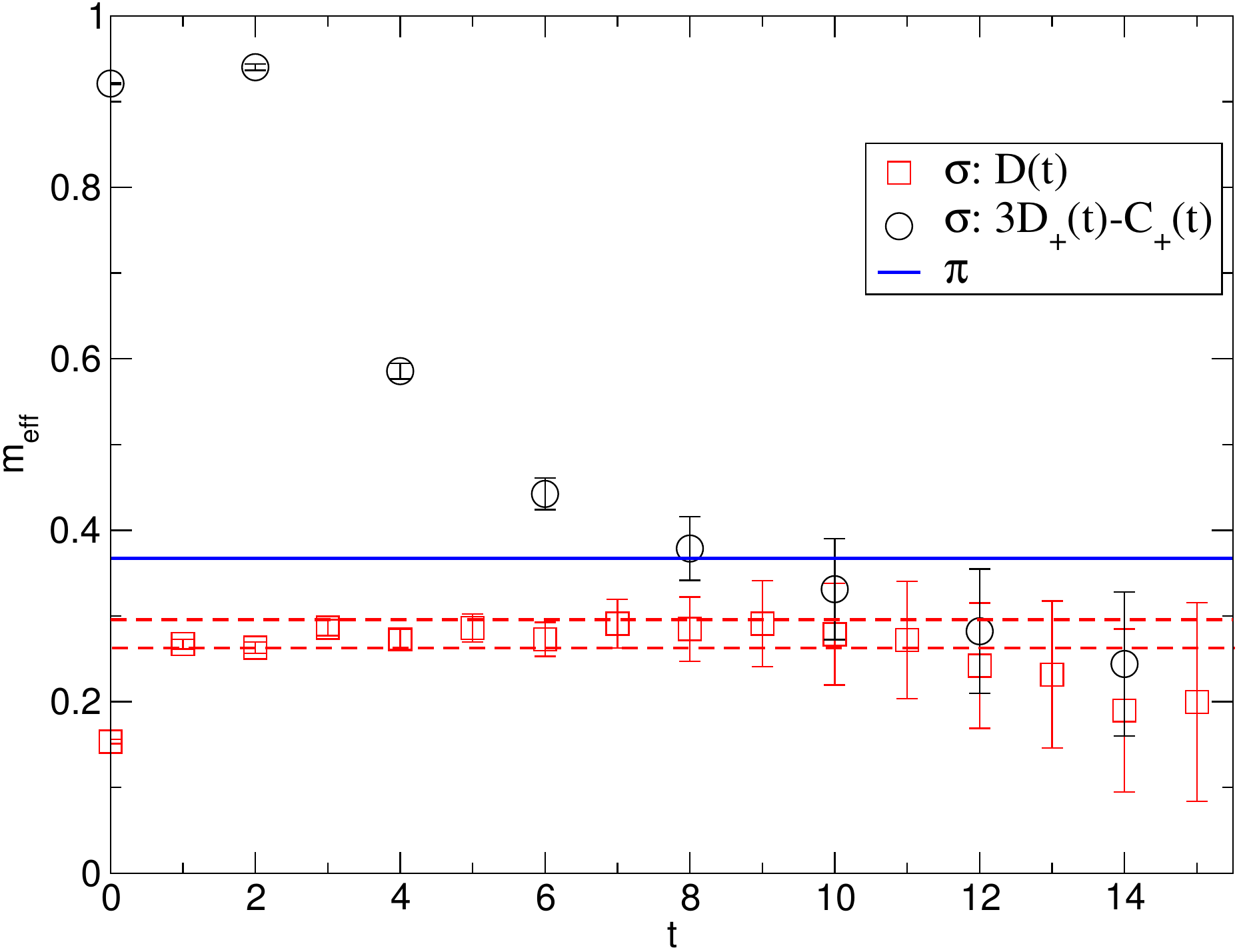}
  \caption{\label{fig:meff_fermionic} Effective scalar mass $m_{\sigma}$ from correlators in Eq.~\eqref{eq:fermionic-combined}, with the projection explained in the text, and in Eq.~\eqref{eq:fermionic-disconnected-only} for $L =24$ and $m_f=0.06$. The dashed and solid lines highlight the fit result for $m_{\sigma}$ with statistical error band and $m_\pi$, respectively.}
\end{figure}

In order to minimize $A_{\pi_{\rm \overline{SC}}}(t)$ in $3D(t)-C(t)$, we adopt a projection, $C_+(t) = 2C(t) + C(t+1) + C(t-1)$, at even $t$. Figure~\ref{fig:meff_fermionic} shows that the effective mass of $3D_+(t)-C_+(t)$ at large $t$ is smaller than $m_\pi$, while the error is large. As an alternative method, we also employ $D(t)$ to extract $m_{\sigma}$, and its effective mass is also shown in the figure. The effective mass plateau of $D(t)$ is consistent with the one of $3D_+(t)-C_+(t)$ in the large--time region. Furthermore the plot clarifies the importance of using $D(t)$ to extract $m_{\sigma}$, because it performs better in identifying the lightest scalar state, even at small temporal separations. This might be caused by a reasonable cancellation among contributions from excited scalar states and the $a_0$ state in $D(t)$. It should be noted that, because of the small $m_\sigma$, the exponential damping of $D(t)$ is slow, which helps preventing the rapid degradation of the signal--to--noise ratio.

We fit $D(t)$ between $t=4$ and $t=8$, assuming a single light state propagating in this region, to obtain $m_\sigma$ for all the parameters. A systematic error coming from the fitting range choice is estimated by the difference of central values obtained with several fit ranges. The results of $m_\sigma$ and $m_\pi$ are reported in Table~\ref{tab:simulations}. We find that $m_{\sigma} < m_{\pi} < m_{a_0}$ for all the investigated fermion masses. The difference of $m_\sigma$ and $m_\pi$ is more than one standard deviation when the statistic
and systematic errors are combined in quadrature, except for $m_f = 0.06$ on $L=30$, where there is a sizable systematic error.

As mentioned before, bound states in the $0^{++}$ channel for a non--abelian strongly--interacting gauge theory can contain gluonic degrees of freedom, as well as fermionic ones. This gluonic content has been studied already in the SU($2$) gauge theory with two adjoint fermions~\cite{DelDebbio:2009fd}. The method to measure glueball masses on the lattice employs a large number of different interpolating operators built from gauge-invariant combinations of gauge links in such a way that a robust basis for a variational ansatz can be created (see for example Refs.~\cite{Morningstar:1997ff,Lucini:2004my,Lucini:2010nv,Gregory:2012hu}). We build gauge--invariant and zero--momentum interpolating operators $\Og^{\alpha}(t)$ with scalar rotational quantum numbers. By using differently shaped spatial Wilson loops, we construct $32$ different basis operators for the scalar glueball. Each of these operators is smeared at several levels ($5$ or $6$) and we obtain a large variational basis $\Og^{\alpha}(t)$, $\alpha=1, \ldots,160$($192$).

\begin{figure}[t]
  \includegraphics[height=5cm]{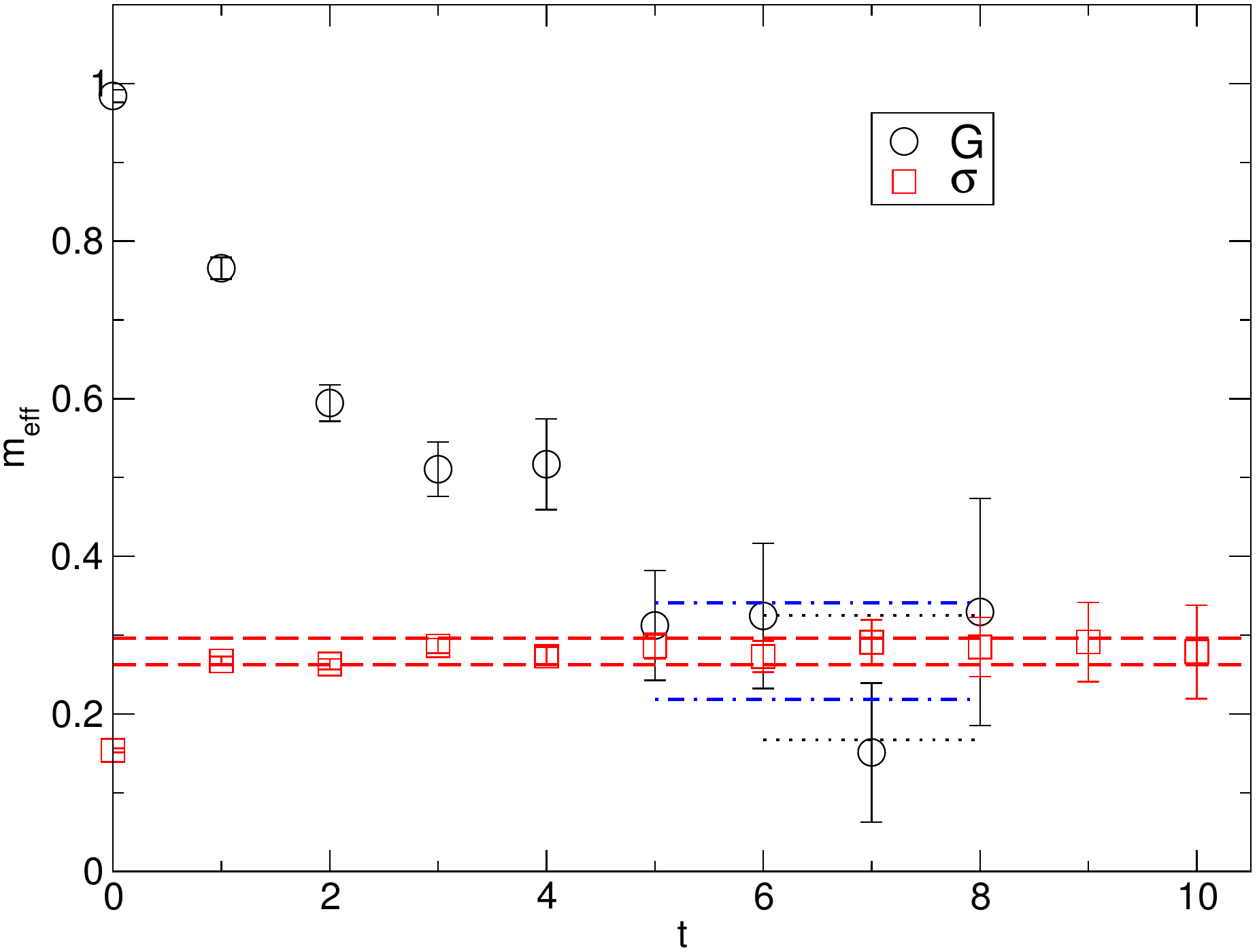}
  \caption{\label{fig:compare-f-g} Fermionic $m_{\sigma}$ and gluonic $m_G$ effective masses (respectively from correlators in Eq.~\eqref{eq:fermionic-disconnected-only} and Eq.~\eqref{eq:gluonic-correlator}) for $L=24$ and $m_f=0.06$. The fitted masses are highlighted by dashed and dotted-dashed lines for the gluonic correlators and dotted lines for the fermionic one. Systematics effects on the gluonic mass are not relevant given the larger statistical error.}
\end{figure}

The variational ansatz proves to be successful in extracting a signal for the ground state from vacuum subtracted cross--correlation matrices of the form 
\begin{equation}
  \label{eq:gluonic-correlator}
   C^{\alpha \beta}(t) \; = \; \langle \Og^{\alpha}(t) \Og^{\beta}(0) \rangle - \langle \Og^{\alpha} \rangle \langle \Og^{\beta} \rangle \ .
\end{equation}
We analyze the correlator after a projection on the eigenstate corresponding to the smallest mass. Figure~\ref{fig:compare-f-g} shows the effective mass of such a state for $m_f=0.06$ on the $L=24$ volume, in comparison with the one obtained from the fermion bilinear (which has already been shown in Fig.~\ref{fig:meff_fermionic}). Remarkably, the asymptotic plateau from both operators agree, though the statistical noise is larger in the gluonic case. The agreement indicates that the gluonic operator has an overlap with the light scalar state which couples to the fermion bilinear. On the $L=24$ volume, we estimate the scalar mass $m_G$ by fitting the large--time behavior ($t=6$--$8$) of the correlator and we obtain $m_G=0.242(68)$ at $m_f=0.05$, $m_G=0.246(79)$ at $m_f=0.06$ and $m_G= 0.28(12)$ at $m_f=0.08$. These $m_G$ are all lighter than $m_\pi$ by more than one standard deviation, while the statistical errors are large.

\begin{figure}[t]
  \includegraphics[height=5cm]{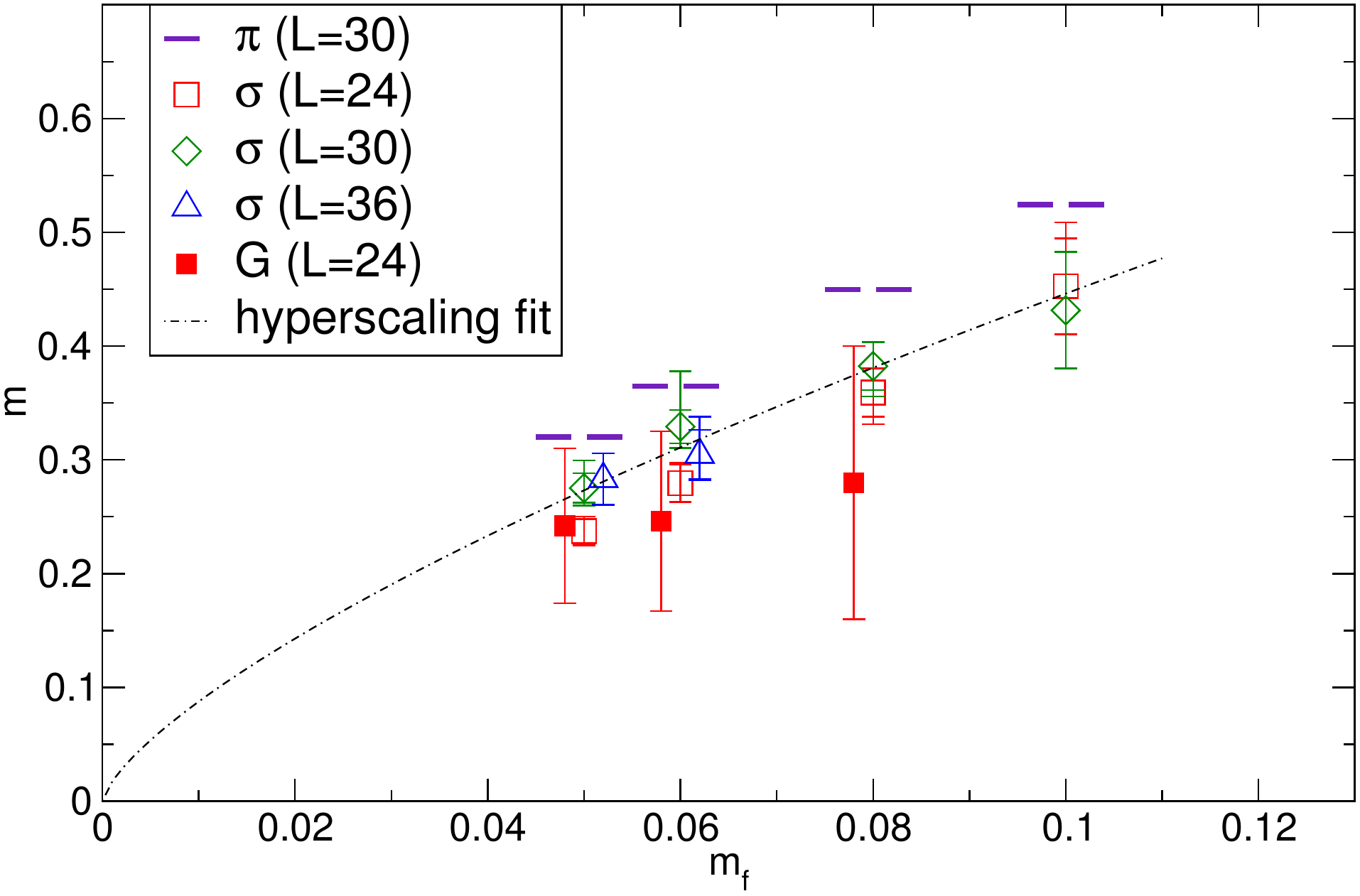}
  \caption{\label{fig:fermionic_spectrum} The mass of the flavor--singlet scalar meson $\sigma$ (see Table~\ref{tab:simulations}) compared to the mass of the pseudo--scalar $\pi$ state and the mass $m_G$ from gluonic operators. Errors are statistical and systematics added in quadrature. The hyperscaling curve is described in the text. The triangle and filleds square symbols are slightly shifted for clarity.}
\end{figure}

Figure~\ref{fig:fermionic_spectrum} presents the flavor--singlet scalar spectrum as function of $m_f$. All the $m_G$'s are consistent with $m_\sigma$ at each parameter. For $m_\sigma$ on the largest two volumes at each $m_f$, finite size effects are negligible in our statistics. For a check of consistency with the hyperscaling of $m_\pi$, we fit $m_\sigma$ on the largest volume data at each $m_f$ using the hyperscaling form $m_{\sigma} = C (m_f)^{1/1+\gamma}$ with a fixed $\gamma = 0.414$ estimated from $m_\pi$~\cite{Aoki:2012eq}, which gives a reasonable value of $\chi^2/{\rm dof}= 0.12$. The fit is shown in Fig.\ref{fig:fermionic_spectrum}. We remind here that the fitted data points have $m_\pi L > 11.5$, as can be checked from Table.~\ref{tab:simulations}. We also estimate the ratio $m_{\sigma}/m_{\pi}$ at each parameter and report it in Table~\ref{tab:simulations}. All the ratios are smaller than unity by more than one standard deviation including the systematic error, except the one at $m_f = 0.06$ on $L=30$, as previously explained. A constant fit with the largest volume data at each $m_f$ gives 0.86(3). These results are consistent with the theory being infrared conformal. Moreover they do not show an abnormal $m_f$ dependence of $m_\sigma$ similar to the one observed in Ref.~\cite{Jin:2012dw}, by which an effect of an unphysical phase boundary would have been suspected.

To summarize, we performed the first study of the scalar flavor--singlet state in $N_f=12$ QCD using fermionic and gluonic interpolating operators. The most striking feature of the measured scalar spectrum is the appearance of a state lighter than the $\pi$ state, as it is shown in Fig.~\ref{fig:fermionic_spectrum}. Such a state appears both in gluonic and fermionic correlators at small bare fermion mass. Clear signals in our simulations were possible thanks to the following salient features: 1. Small taste--symmetry breaking, 2. Efficient noise--reduction methods, 3. Large configuration ensembles, and 4. Slow damping of $D(t)$ thanks to small $m_\sigma$.

We regard the light scalar state observed for $N_f=12$ in this study, as a reflection of the dilatonic nature of the conformal dynamics, since otherwise the $p$--wave bound state (scalar) is expected to be heavier than the $s$--wave one (pseudo--scalar). Thus, it is a promising signal for a walking theory, where a similar conformal dynamics in a wide infrared region should be operative in the chiral limit to form a dilatonic state with mass of $\mathcal{O}(F_\pi)$, in such a way that the tiny spontaneous--breaking--scale $F_\pi$ plays the role of $m_f$ (cfr. Ref.~\cite{Aoki:2013xza}).

While further investigation of the scalar state in $N_f = 12$ QCD, such as a possible lattice spacing dependence, is important, the most pressing future direction is to look at more viable candidates for walking technicolor models. For example, it will be interesting to investigate the scalar spectrum of the $N_f=8$ SU($3$) theory, which was shown to be a good candidate for the walking technicolor~\cite{Aoki:2013xza}, where the scalar state could be identified with the technidilaton, a pseudo Nambu--Goldstone boson coming from the dynamical breaking of conformal symmetry. There actually exists an indication of such a light scalar in $N_f=8$ QCD~\cite{AokiPrep}.

\noindent {\it Acknowledgments.--}
Numerical simulation has been carried out on the supercomputer system $\varphi$ at KMI in Nagoya University, and the computer facilities of the Research Institute for Information Technology in Kyushu University. This work is supported by the JSPS Grant-in-Aid for Scientific Research (S) No.22224003, (C) No.23540300 (K.Y.), for Young Scientists (B) No.25800139 (H.O.) and No.25800138 (T.Y.), and also by Grants-in-Aid of the Japanese Ministry for Scientific Research on Innovative Areas No.23105708 (T.Y.). E.R. was supported by a SUPA Prize Studentship and a FY2012 JSPS Postdoctoral Fellowship for Foreign Researchers (short-term). We would like to thank Luigi~Del~Debbio and Julius~Kuti for fruitful discussions.

\bibliography{reference}

\end{document}